# Direct current generation due to wave mixing in zigzag carbon nanotubes

S. S. Abukari[a], S.Y.Mensah[a], N. G. Mensah[b], K. A. Dompreh[a], A. Twum[a] and F. K. A. Allotey[c]

[a]Department of Physics, Laser and Fibre Optics Centre, University of Cape Coast, Cape Coast, Ghana

[b]Department of Mathematics, University of Cape Coast, Cape Coast, Ghana

[c]Institute of Mathematical Sciences, Accra, Ghana

[*]*Corresponding author.* [a]Department of Physics, Laser and Fibre Optics Centre, University of Cape Coast, Cape Coast, Ghana

Tel.:+233 042 33837

E-mail address: profsymensah@yahoo.co.uk

(S. Y. Mensah)

**Abstract**

Generation of direct current in zigzag carbon nanotubes due to harmonic mixing of two coherent electromagnetic waves is being considered. The electromagnetic waves have commensurate frequencies of $\omega_1 = \Omega$ and $\omega_2 = 2\Omega$. The rectification of the waves at high frequencies. i.e. $\Omega\tau \gg 1$ is quite smooth whiles at low frequencies there are some fluctuations. The nonohmicity observed in the $I - V$ characteristics is attributed to the nonparabolicity of the electron energy band which is very strong in carbon nanotubes because of high stark component. It is observed that the current falls off faster at lower electric field than the case in superlattice. For $\Omega\tau = 2$, the external electric field strength $E^{max}$ for the observation of negative differential conductivity occurs around $1.03 \times 10^6 \, V/m$ which is quite weak. It is interesting to note that the peak of the curve shifts to the left with increasing value of $\Omega\tau$.

1. **Introduction**

It is a known fact that coherent mixing of waves with commensurate frequencies in a nonlinear medium can result in a product which has a zero frequency or static (d.c) electromagnetic field. If such a nonlinear interference phenomenon happens in a semiconductor or semiconductor device, then the static electric field may result into a d.c current or a dc voltage generation [1].

Infact, several mechanisms of nonlinearity could be responsible for the wave mixing in semiconductors [2-4]. Important among them is the heating mechanism where the nonlinearity is related to the dependence of the relaxation constant on the electric field [4-7] and the nonparabolicity of electron energy spectrum [1]. Goychuk and Hänggi [8] have also suggested another scheme of quantum rectification using wave mixing of an alternating electric field and its second harmonic in a single miniband superlattice (SL). Their approach is based on the theory of quantum ratchets and therefore the necessary conditions for the appearance of dc include a dissipation (quantum noise) and an extended periodic system [8].

Interesting to this paper is where the mechanism of nonlinearity is due to the nonparabolocity of the electron energy spectrum. Notable among such materials are the superlattice (SL) and carbon nanotubes (CNs). In superlattice the theory of wave mixing based on a solution of the Boltzmann equation have been studied in [9-11]. In all these works, the situation where $E(t) = E_1 cos\Omega t + E_2 cos(2\Omega t + \varphi)$ were not studied directly. The first paper to study this situation in SL can be found in [12]. Recently this problem has been revisited in the following papers [1,13, 14] because of the interest it generates. We study this effect in zigzag carbon nanotubes.

This work will be organised as follows: section 1 deals with introduction; in section 2, we establish the theory and solution of the problem; section 3, we discuss the results and draw conclusion.

## 2. Theory

Following the approach of [15] we consider an undoped single-wall zigzag (n, 0) carbon nanotubes (CNs) subjected to the electric mixing harmonic fields.

$$E(t) = E_1 cos\omega_1 t + E_2 cos(\omega_2 t + \theta) \qquad (1)$$

We further consider the semiclassical approximation in which the motion of $\pi$-electrons are considered as classical motion of free quasi-particles in the field of crystalline lattice with dispersion law extracted from the quantum theory.

Considering the hexagonal crystalline structure of CNs and the tight binding approximation, the dispersion relation is given as

$$\begin{aligned}\varepsilon(s\Delta p_\varphi, p_z) &\equiv \varepsilon_s(p_z) \\ &= \pm\gamma_0 \left[1 + 4cos(ap_z)cos\left(\frac{a}{\sqrt{3}}s\Delta p_\varphi\right) \right. \\ &\left. + 4cos^2\left(\frac{a}{\sqrt{3}}s\Delta p_\varphi\right)\right]^{1/2}\end{aligned} \qquad (2)$$

for zigzag CNs [15]

Where $\gamma_0 \sim 3.0eV$ is the overlapping integral, $p_z$ is the axial component of quasimomentum, $\Delta p_\varphi$ is transverse quasimomentum level spacing and $s$ is an integer. The expression for $a$ in Eq (2) is given as

$$a = {3a_{c-c}}/{2\hbar} \qquad (3)$$

with the C-C bond length $a_{c-c} = 0.142nm$ and the Plank's constant $\hbar$, $h/2\pi$. The - and + signs correspond to the valence and conduction bands, respectively. Due to the transverse quantization of the quasi-momentum, its transverse component can take $n$ discrete values, $p_\varphi = s\Delta p_\varphi = \pi\sqrt{3}\ \pi/an\ (s = 1 ...., n)$

Unlike transverse quasimomentum $p_\varphi$, the axial quasimomentum $p_z$ is assumed to vary continuously within the range $0 \leq p_z \leq 2\pi/a$, which corresponds to the model of infinitely long $CN(L = \infty)$. This model is applicable to the case under consideration because of the restriction to the temperatures and /or voltages well above the level spacing [16], ie. $k_B T > \varepsilon_C, \Delta\varepsilon$, where $k_B$ is Boltzmann constant, $T$ is the temperature, $\varepsilon_C$ is the charging energy. The energy level spacing $\Delta\varepsilon$ is given by

$$\Delta\varepsilon = \pi\hbar v_F / L \qquad (4)$$

where $v_F$ is the Fermi velocity and $L$ is the carbon nanotube length [17]

Employing Boltzmann equation with a single relaxation time approximation

$$\frac{\partial f(p)}{\partial t} + eE(t)\frac{\partial f(p)}{\partial P} = -\frac{[f(p) - f_0(p)]}{\tau} \qquad (5)$$

where $e$ is the electron charge, $f_0(p)$ is the equilibrium distribution function, $f(p,t)$ is the distribution function, and $\tau$ is the relaxation time. The electric field $E$ is applied along CNs axis. In this problem the relaxation term $\tau$ is assumed to be constant. The justification for $\tau$ being constant can be found in [18]. The relaxation term of Eq. (5) describes the effects of the dominant type of scattering (e.g. electron-phonon and electron-twistons) [19]. For the electron scattering by twistons (thermally activated twist deformations of the tube lattice), $\tau$ is proportional to $m$ and the $I - V$ characteristics have shown that scattering by twistons increases $E^{max}$ and decreases $|\partial j_z/\partial E_z|$ in the NDC region; the lesser $m$, the stronger this effect. Quantitative changes of the $I - V$ curves turn out to be insignificant in comparison with the case of $\tau = $ const [18, 19].

Expanding the distribution functions of interest in Fourier series as;

$$f_0(p) = \Delta p_\varphi \sum_{s=1}^{n} \delta(p_\varphi - s\Delta p_\varphi) \sum_{r \neq 0} f_{rs}\, e^{iarp_z} \qquad (6)$$

and

$$f(p,t) = \Delta p_\varphi \sum_{s=1}^{n} \delta(p_\varphi - s\Delta p_\varphi) \sum_{r \neq 0} f_{rs}\, e^{iarp_z} \emptyset_\upsilon(t) \qquad (7)$$

Where the coefficient, $\delta(x)$ is the Dirac delta function, $f_{rs}$ is the coefficient of the Fourier series and $\emptyset_\upsilon(t)$ is the factor by which the Fourier transform of the nonequilibrium distribution function differs from its equilibrium distribution counterpart.

$$f_{rs} = \frac{a}{2\pi\Delta p_\varphi S}\int_0^{\frac{2\pi}{a}} \frac{e^{-iarp_z}}{1 + \exp(\varepsilon_s(p_z)/k_B T)}\, dp_z \qquad (8)$$

Substituting Eqs. (6) and (7) into Eq. (5), and solving with Eq. (1) we obtain

$$\emptyset_v(t) = \sum_{k_1,k_2=-\infty}^{\infty} \sum_{v_1,v_2=-\infty}^{\infty} J_{k_1}(r\beta_1) J_{k_2}(r\beta_2) J_{k_1+v_1}(r\beta_1) J_{k_2+v_2}(r\beta_2)$$

$$\times \left( \frac{(1 - i(k_1\omega_1 + k_2\omega_2)\tau)}{1 + ((k_1\omega_1 + k_2\omega_2)\tau)^2} \right) \times \{\cos(v_1\omega_1 t + v_2(\omega_2 t + \theta))$$

$$- i\sin(v_1\omega_1 t + v_2(\omega_2 t + \theta))\} \quad (9)$$

where $\beta_1 = \frac{eaE_1}{\omega_1}$, $\beta_2 = \frac{eaE_2}{\omega_2}$, and $J_k(\beta)$ is the Bessel function of the $k^{th}$ order.

Similarly, expanding $\varepsilon_s(p_z)/\gamma_0$ in Fourier series with coefficients $\varepsilon_{rs}$

$$\frac{\varepsilon_s(p_s, s\Delta p_\varphi)}{\gamma_0} = \varepsilon_s(p_z) = \sum_{r \neq 0} \varepsilon_{rs} e^{iearp_z} \quad (10)$$

Where $\varepsilon_{rs} = \frac{a}{2\pi\gamma_0} \int_0^{\frac{2\pi}{a}} \varepsilon_s(p_z) e^{-iearp_z} dp_z$ \quad (11)

and expressing the velocity as

$$v_z(p_z, s\Delta p_\varphi) = \frac{\partial \varepsilon_s(p_z)}{\partial p_z} = \gamma_0 \sum_{r \neq 0} iar \, \varepsilon_{rs} e^{iearp_z} \quad (12)$$

We determine the surface current density as

$$j_z = \frac{2e}{(2\pi\hbar)^2} \iint f(p) \, v_z(p) d^2p,$$

or

$$j_z = \frac{2e}{(2\pi\hbar)^2} \sum_{s=1}^{n} \int_0^{\frac{2\pi}{a}} f\left(p_z, s\Delta p_\varphi, \emptyset_v(t)\right) v_z(p_z, s\Delta p_\varphi) dp_z \quad (13)$$

and the integration is taken over the first Brillouin zone. Substituting Eqs. (7), (9) and (12) into (13) and linearizing with respect to $E_2$ using $J_{\pm 1}(\beta_2) \sim \beta_2/2$ ; $J_0(\beta_2) \sim 1 - \left(\beta_2^2/4\right)$ and then averaging the result with respect to time $t$, we obtain the DC subjected to $\omega_1 = \Omega$ and $\omega_2 = 2\Omega$ as follows;

$$j_z = \frac{2e^2\gamma_0 a}{\sqrt{3}\hbar n a_{c-c}} E_2 \cos\theta \sum_{r=1}^{\infty} r^2 \sum_{k=-\infty}^{\infty} \frac{k J_k(r\beta_1) J_{k-2}(r\beta_1)}{1 + (k\Omega\tau)^2} \sum_{s=1}^{n} f_{rs}\varepsilon_{rs} \quad (14)$$

## 3. Results, Discussion and Conclusion

Using the solution of the Boltzmann equation with constant relaxation time τ, the exact expression for current density in CNs subjected to an electric field with two frequencies $\omega_1 = \Omega$ and $\omega_2 = 2\Omega$ was obtained.

We noted that the current density $j_z$ is dependent on the electric field $E_2$ and $E_1$, the phase difference $\theta$, the frequency $\Omega$, the relaxation time $\tau$ and $n$. To further understand how these parameters affect $j_z$, we sketched equation (14) using Matlab. Fig.1 represents the graph of $j_z/j_o$ on $\beta_1$ for $\Omega\tau = 0.3, 0.5, 0.9, 1\ and\ 2$. We observed that as $\beta_1$ increases, the magnitude of the normalized current $j_z/j_o$ increases and at $\beta_1 = \beta_1^{max}$ the normalized current reaches a maximum value. Further increase of $\beta_1$ results in the decrease of the magnitude of $j_z/j_o$. Thus, the region with negative differential conductivity (NDC) $\partial j_z/\partial \beta_1 < 0$. We further observed that for $r = 1$, the curve for the nanotube is qualitatively the same as the superlattice except that the peaks occur at higher values see Fig.1(a) and 2. As we increase the $r$ values the magnitude of the normalized current also increases. For $r > 1$ we observed some fluctuations in the NDC region when $\Omega\tau \ll 1$. On the other hand for $\Omega\tau \gg 1$ we observed that the current decreases monotonically in the NDC region. This indicates that at high frequency the rectification is smooth as against the case of low frequency. See Figs. 1(b)-(e). The rectification can be attributed to non ohmicity of the carbon nanotube for the situation where it Bloch oscillates and the high stark component (summation over $r$). From Fig. 1(d) for example, if $\Omega\tau = 2$, the external electric field strength $E^{max}$ necessary to induce the stark frequency for the observation of negative differential conductivity occurs around $1.03 \times 10^6\ V/m$ which is quite weak. As can be seen from Figs. 1(b)-(e) the peak of the current shifts to the left with increasing value of $\Omega\tau$. In comparing the peak values of Fig. 1(d) with the result in [12] (see Fig.2) for $\Omega\tau = 1$, the ratio $\frac{|j_{max}^{CNs}|}{|j_{max}^{SL}|} \approx 40$ and for or $\Omega\tau = 2$, the ratio $\frac{|j_{max}^{CNs}|}{|j_{max}^{SL}|} \approx 44$ which is quite substantial.

We sketched also the graph of $j_z/j_o$ against $\Omega\tau$ for $\beta_1 = 0.3, 0.5, 0.9, 1\ and\ 2$. The graph also displayed a NDC. See Fig.3. Interestingly unlike in SL as indicated in [1] where the current is positive and has a maximum at the value $\Omega\tau \approx 0.71$ irrespective of the amplitude of the electric field $E_1$ in carbon nanotubes because of the high stark component the maximum peak shifts towards the left with increasing value of $\beta_1$ see Figs. 3 (b)-(d). However, when $r = 1$ as stated above it behaves like SL. See Fig. 3(a). It is worthwhile to note that $\Omega\tau_{max}$ can be used to determine the relaxation time of the dominant type of scattering (i.e. electron-phonon) in the nanotube. It is important to note that when the phase shifts $\theta$ lies between $\frac{\pi}{2}$ and $\frac{3\pi}{2}$ there is an inversion. See Fig. 4.

In conclusion, we have studied the direct current generation due to the harmonic wave mixing in zigzag carbon nanotubes and suggest the use of this approach in generation of THz radiation. The experimental conditions for an observation of the dc current effect are practically identical to those fulfilled in a recent experiment on the generation of harmonics of the THz radiation in a semiconductor superlattice [20]. This method can also be used to determine the relaxation time $\tau$.

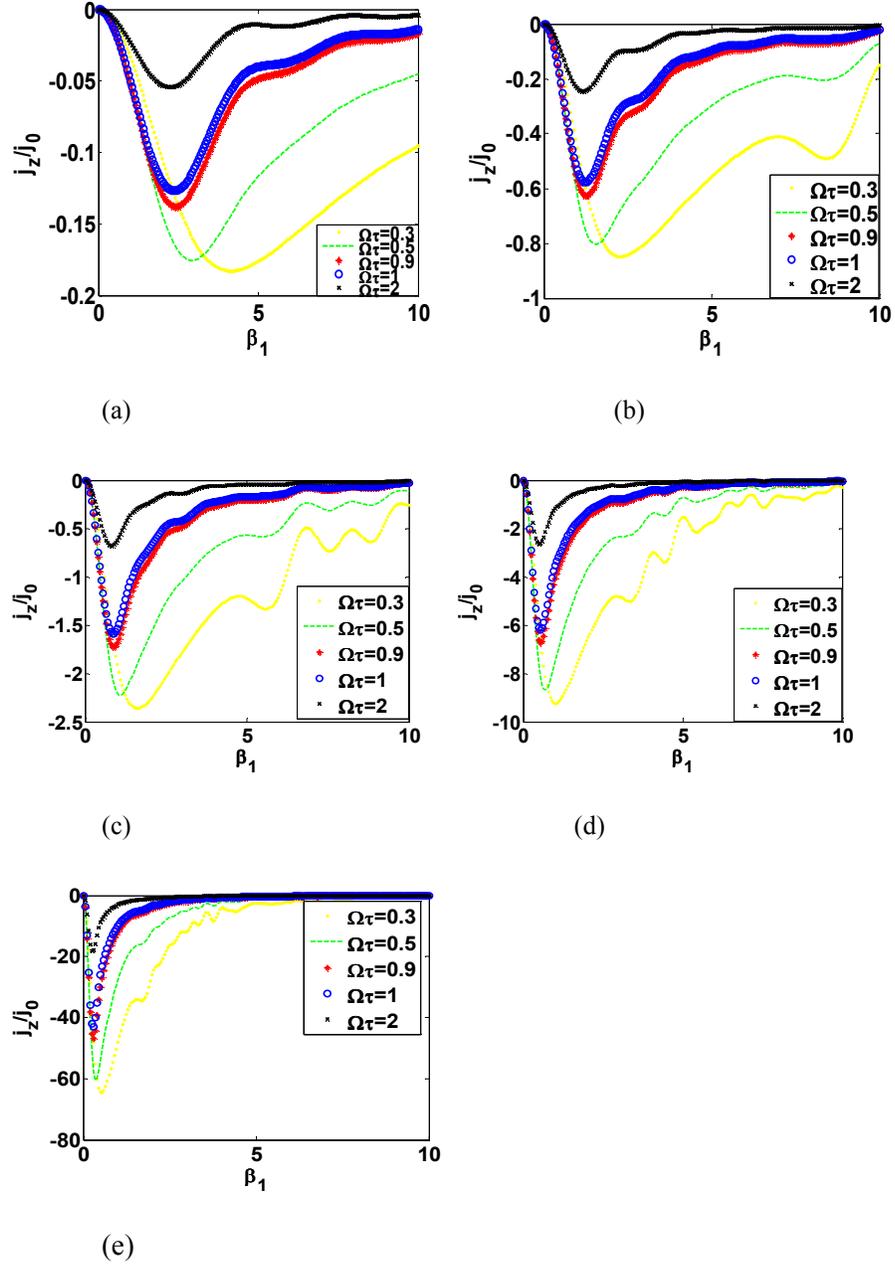

**Fig. 1**. $j_z/j_o$ is plotted against $\beta_1$ for (⋯) $\Omega\tau = 0.3$; (⋯) $\Omega\tau = 0.5$; (∗∗∗∗) $\Omega\tau = 0.9$ ($oooo$) $\Omega\tau = 1$; $and$ (⋯) $\Omega\tau = 2$. When (a) $r = 1$, (b) $r = 2$, (c) $r = 3$, (d) $r = 5$, and (e) $r = 10$.

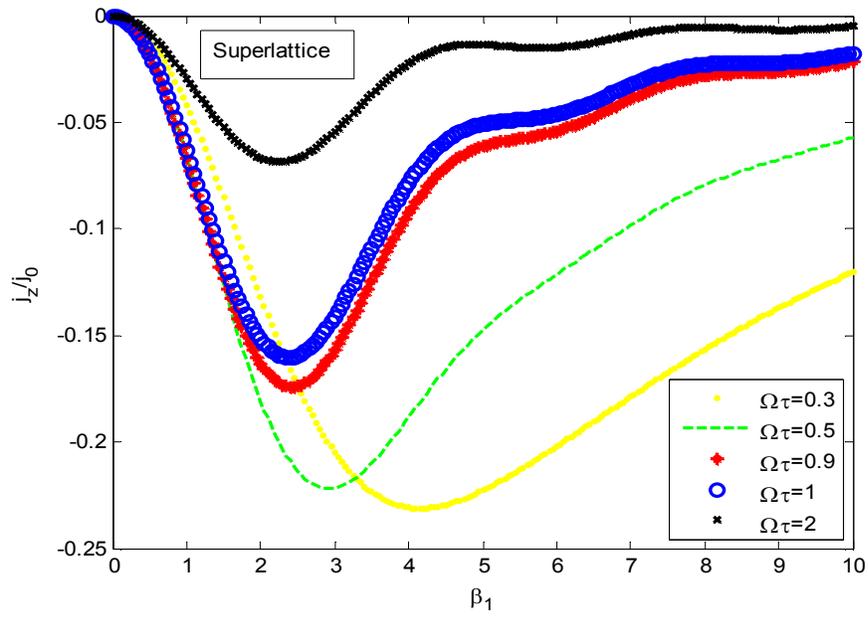

**Fig. 2.** . $j_z/j_o$ is plotted against $\beta_1$ for (⋯) $\Omega\tau = 0.3$; (⋯) $\Omega\tau = 0.5$; (∗∗∗∗) $\Omega\tau = 0.9$ (oooo)$\Omega\tau = 1$; and (⋯) $\Omega\tau = 2$.

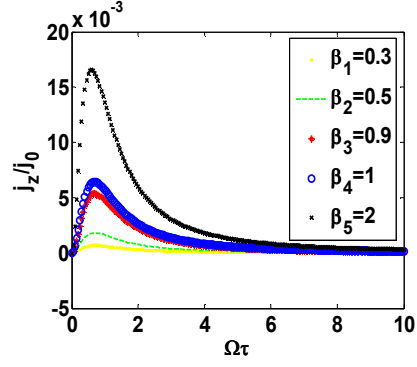
(a)

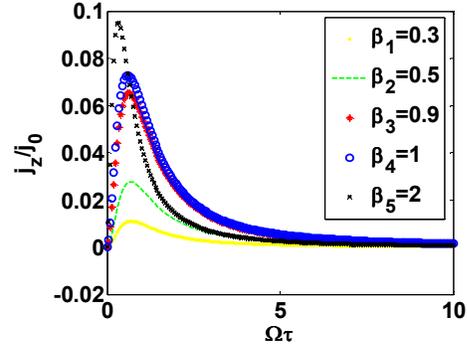
(b)

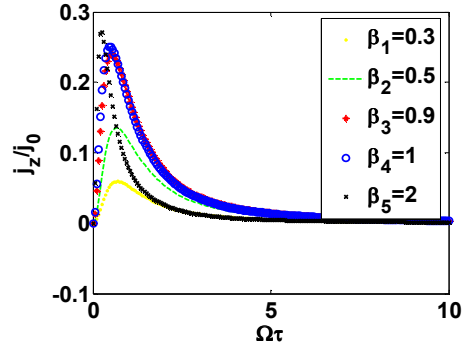
(c)

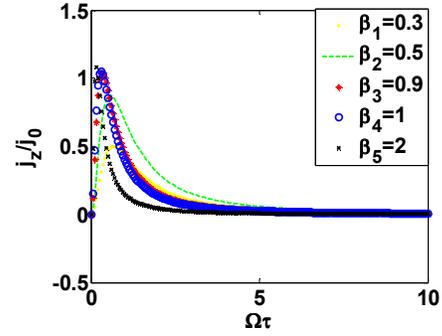
(d)

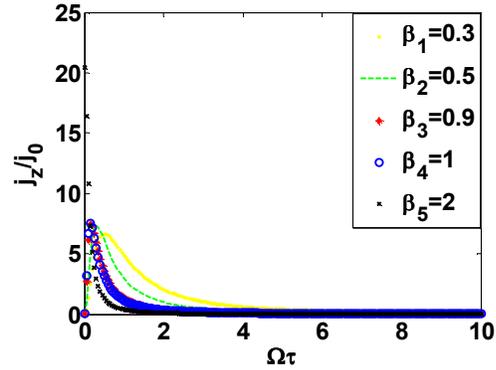
(e)

**Fig. 3.** $j_z/j_o$ is plotted against $\Omega\tau$ for (· — ·) $\beta_1 = 0.3$; (···) $\beta_1 = 0.5$; (***) $\beta_1 = 0.9$ ; (ooo) $\beta_1 = 1$; and (···) $\beta_1 = 2$. When (a) $r = 1$, (b) $r = 2$, (c) $r = 3$, (d) $r = 5$, and (e) $r = 10$.

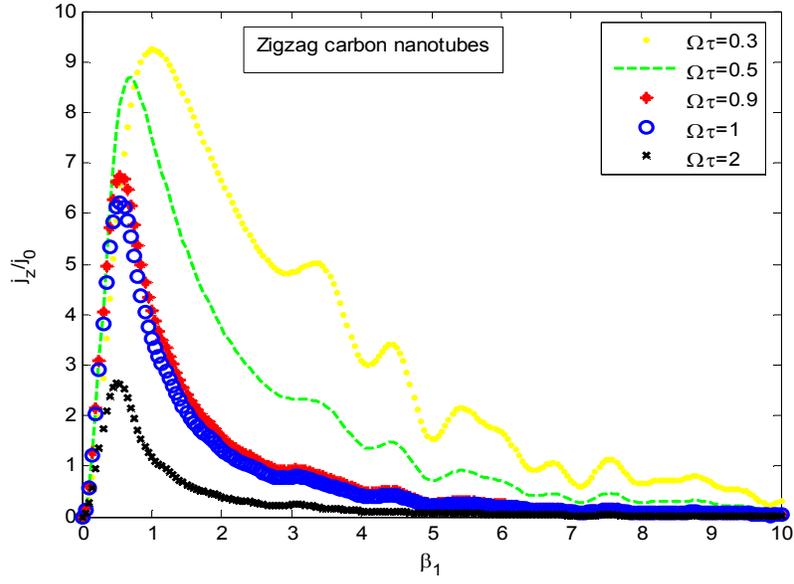

**Fig. 4.** $j_z/j_o$ is plotted against $\beta_1$ for $(\cdots)$ $\Omega\tau = 0.3$; $(\cdots)$ $\Omega\tau = 0.5$; $(****)$ $\Omega\tau = 0.9$ $(oooo)$ $\Omega\tau = 1$ $and$ $(\cdots)$ $\Omega\tau = 2$. When $r = 5$ and the phase shift $\theta$ lies between $\frac{\pi}{2}$ and $\frac{3\pi}{2}$.

**References**

[1] Kiril N. Alekseev and Feodor V. Kusmartsev  arxiv: cond-mat/0012348v1, 2000

[2] Patel C. K. N., Slusher R.E. and Fleury P.A., Phys Rev Lett., 17 (1966) 1011

[3] Wolf P.A. and Pearson G.A., Phys. Rev. Lett., 17 (1966) 1015

[4] Belyantsev A. M., Kozlov V. A. and Trifonov B.A, Phys. Status Sol. (b) 48 (1971) 581

[5] Fomin V. M. and Pokatilov E.P., Phys. Status Sol. (b) 97 (1980) 161

[6] Shmelev G, M., Tsurkan G. J. and Nguyen Xong Shon, Izv. Vyzov, Fizika 2(1985) 84 (Russian Physics Journal)

[7] Genkin V. N., Kozlov V. A. and Piskarev V.I. Fiz. Tekh. Polupr., 8 (1074) 2013 (Sov. Phys. Semicond)

[8] Goychuk I., and Hänggi P., Euophys. Lett., 43 (1998) 503

[9] Romanov Yu A., Orlov L. K. and Bovin V.P., Fiz. Tekhn. Polupr. (1978) 1665

[10] Orlov L. K. and Romanov Yu A., Fiz. Tverd. Tela 19 (1977) 726

[11] Pavlovich V.V., Fiz. Tverd. Tela 19 (1977) 97

[12] Mensah S. Shmelev G. M. and Epshtein E.M., Izv. Vyzov, Fizika 6 (1988) 112

[13] Kirill N. Alekssev, Mikhael V. Erementchouk and Feodor V. Kusmarttsev, con-mat/9903092v1 1999

[14] Seeger K. Applied Physics Letters. Vol 76 No1 (2000)

[15] Anton S. Maksimenko and Gregory Ya. Slepan, Physical Review Letters. Vol 84 No 2 (2000)

[16] Kane, C., Balents, L. and Fisher, M. P. A., Phys. Rev. Lett. 79, 5086–5089, 1997.

[17] Lin, M. F. and Shung, K. W.K., Phys. Rev. B 52, pp. 8423–8438, 1995.

[18] Kane, C. L., Mele, E. J., Lee, R. S., Fischer, J. E., Petit, P., Dai, H., Thess, A., Smalley, R. E., Verscheueren, A. R. M., Tans, S. J. and Dekker, C., Europhys. Lett. 41, 683-688 (1998).

[19] Jishi, R. A., Dresselhaus, M. S., and Dresselhaus, G. Phys. Rev. B 48, 11385 - 11389, (1993).

[20] Winnerl, S. Schomburg, E. Brandl, S. Kus, O., Renk, K. F.. Wanke, M. C. Allen, S. J, Ignatov, A. Ustinov, V. Zhukov, A. and. Kop'ev, P. S.  Appl. Phys. Lett. 77, 1259 (2000).